\documentclass[twocolumn,showpacs,preprintnumbers,amsmath,amssymb,epsf,prl]{revtex4}

\usepackage{graphicx}% Include figure files
\usepackage{dcolumn}% Align table columns on decimal point
\usepackage{bm}% bold math
\usepackage{epsfig}

\begin{document}
%\documentstyle[aps,prl,twocolumn,epsf]{revtex}
%\begin{document}
%\draft
%\wideabs{
\title{Critical rotation of a harmonically trapped Bose gas}
\author{P. Rosenbusch$^1$, D.S. Petrov$^{2,3}$, S. Sinha$^1$, F. Chevy$^1$,
V. Bretin$^1$, Y. Castin$^1$, G. Shlyapnikov$^{1,2,3}$, and J. Dalibard$^1$}
\affiliation{$^1$ Laboratoire Kastler Brossel$^*$, 24 rue Lhomond,
75005 Paris, France\\
$^{2}$ FOM Institute for Atomic and Molecular Physics,
Kruislaan 407, 1098 SJ Amsterdam, The Netherlands \\
$^{3}$ Russian Research Center Kurchatov Institute, Kurchatov square,
123182 Moscow, Russia}
\date{Received }

%\maketitle

\begin{abstract}
{We study experimentally and theoretically a cold trapped Bose gas
under critical rotation, \emph{i.e.} with a rotation frequency
close to the frequency of the radial confinement. We identify two
regimes: the regime of explosion where the cloud expands to infinity in
one direction, and the regime where the condensate spirals out of
the trap as a rigid body. The former is realized for a dilute
cloud, and the latter for a Bose-Einstein condensate with the
interparticle interaction exceeding a critical value. This
constitutes a novel system in which repulsive interactions help in
maintaining particles together.}
\end{abstract}

\pacs{03.75.Fi, 32.80.Lg}
%\narrowtext

\maketitle

The rotation of a macroscopic quantum object is a source of
spectacular and counter-intuitive phenomena. In superfluid liquid
helium contained in a cylindrical bucket rotating around its axis
$z$, one observes the nucleation of quantized vortices for a
sufficiently large rotation frequency $\Omega$ \cite{Donnelly}. A
similar phenomenon occurs in a Bose-Einstein condensate confined
in a rotating harmonic trap \cite{ENS,MIT,Boulder,Oxford}. In
particular, vortices are nucleated when the rotation resonantly
excites surface modes of the condensate. This occurs for
particular rotation frequencies in the interval $0<\Omega \leq
\omega_\bot/\sqrt{2}$, where $\omega_\bot$ is the trap frequency
in the $xy$ plane perpendicular to the rotation axis $z$.

Several theoretical studies have recently considered the critical
rotation of the gas, \emph{i.e.} $\Omega\sim \omega_\bot$, which
presents remarkable features
\cite{Rokhsar,Mottelson,Gunn,Stringari,Zoller,Ho,Fetter,Baym,Sinova}.
 From a classical point of view, for $\Omega=\omega_\bot$ the
centrifugal force compensates the harmonic trapping force
in the $xy$ plane. Hence the motion of a single particle of mass
$m$ in the frame rotating at frequency $\Omega$ is simply due to
the Coriolis force $2 m  {\bf{\dot r}}\times {\bf \Omega} $. This
force is identical to the Lorentz force acting on a particle of
charge $q$ in the magnetic field ${\bf B}= 2 (m/q)\, {\bf
\Omega}$. The analogy between the motion of charged
particles in a magnetic field and neutral particles in a rotating
frame also holds in quantum mechanics. In this respect, a quantum
gas of atoms confined in a harmonic trap rotating at the critical
frequency is analogous to an electron gas in a uniform
magnetic field. One can then expect
\cite{Gunn,Zoller} to observe phenomena related
to the Quantum Hall Effect.

This paper presents an experimental and theoretical study of the
dynamics of a magnetically trapped rubidium ($^{87}$Rb) gas
stirred at a frequency close to $\omega_\bot$. We show that the
single particle motion is dynamically unstable for a window of
frequencies $\Omega$ centered around $\omega_\bot$. This result
entails that the center-of-mass of the atom cloud (without or with
interatomic interactions) is destabilized, since its motion is
decoupled from any other degree of freedom for a harmonic
confinement. This also implies that a gas of non-interacting
particles ``explodes'', which we indeed check experimentally. When
one takes into account the repulsive interactions between
particles, which play an important role in a $^{87}$Rb condensate,
one would expect naively that this explosion is enhanced. However,
we show experimentally that this is not the case, and repulsive
interactions can ``maintain the atoms together". This has been
predicted for a Bose-Einstein condensate in the strongly
interacting -- Thomas-Fermi (TF)-- regime \cite{Stringari}. Here
we derive the minimal interaction strength which is necessary to
prevent the explosion. This should help studies of the Quantum
Hall related physics in the region of critical rotation.

Consider a gas of particles confined in an axisymmetric harmonic
potential $V_0({\bf r})$, with frequency $\omega_z$ along the trap
axis $z$, and $\omega_\bot$ in the $xy$ plane. To set this gas
into rotation, one superimposes a rotating asymmetric potential in
the $xy$ plane. In the reference frame rotating at an angular
frequency $\Omega$ around the $z$ axis, this potential reads
$V_1({\bf r})=\epsilon m\omega_\bot^2 (Y^2 -X^2)/2$, where
$\epsilon>0$. The rotating frame coordinates $X,Y$ are deduced
from the lab frame coordinates $x,y$ by a rotation at an angle
$\Omega t$.

For a non-interacting gas, the equation of motion for each
particle reads:
\begin{eqnarray}
& & \ddot X -2 \Omega \dot
Y+\left(\omega_\bot^2(1-\epsilon)-\Omega^2
\right)X=0 \label{eq:Xmotion}\\
& & \ddot Y +2 \Omega \dot
X+\left(\omega_\bot^2(1+\epsilon)-\Omega^2 \right)Y=0,
\label{eq:Ymotion}
\end{eqnarray}
while the motion along $z$ is not affected by the rotation. One
deduces from this set of equations that the motion in the $xy$
plane is dynamically unstable if the stirring frequency $\Omega$
is in the interval $[\omega_\bot \;\sqrt{1-\epsilon},\omega_\bot\;
\sqrt{1+\epsilon}]$. In particular, for $\Omega=\omega_\bot$ and
$\epsilon \ll 1$, one finds that the quantity $X+Y$ diverges as
$\exp{(\epsilon \omega_\bot t/2)}$, whereas $X-Y$ remains finite.
%The gas extension along the $X=Y$
%line becomes very large, so that
%the atomic cloud looks after some
%time like the rotating blade of
%a propeller.

To test this prediction we use a $^{87}$Rb cold gas in a
Ioffe-Pritchard magnetic trap, with frequencies $\omega_x=
\omega_y=2\pi\times 180$~Hz, and $\omega_z=2\pi \times 11.7$~Hz.
The initial temperature of the cloud pre-cooled using optical
molasses is 100 $\mu$K. The gas is further cooled by
radio-frequency evaporation. For the first set of experiments we
stop the evaporation before the Bose-Einstein condensation is
reached. The resulting sample contains $10^7$ atoms at a
temperature $T\sim 5\,\mu$K. It is dilute, with a central density
$\sim 10^{12}$~cm$^{-3}$, and atomic interactions can be neglected
(mean-field energy $\ll k_B T$). The second set of experiments
corresponds to a much colder sample ($T<50$~nK), \emph{i.e.} to a
quasi-pure condensate with $10^5$ atoms.

After evaporative cooling, the atomic cloud is stirred during an
adjustable period $t$ by a focused laser beam of wavelength
$852$~nm and waist $w_0=20 \mu$m, whose position is controlled
using acousto-optic modulators \cite{ENS}. The beam is switched on
abruptly and it creates a rotating optical-dipole potential which
is nearly harmonic over the extension of the cloud. We measure the
transverse density profile of the condensate after a period of
free expansion. In this pursuit, we suddenly switch off the
magnetic field and the stirrer, allow for a 25~ms free-fall, and
image the absorption of a resonant laser by the expanded cloud.
The imaging beam propagates along the $z$ axis. We fit the density
profile of the sample assuming a Gaussian shape for the
non-condensed cloud, and a parabolic TF shape for the quasi-pure
condensate. We extract from the fit the long and short diameters
in the plane $z=0$, and the average position of the cloud. The
latter gives access to the velocity of the center-of-mass of the
atom cloud before time of flight.

\begin{figure}
\centerline{\includegraphics{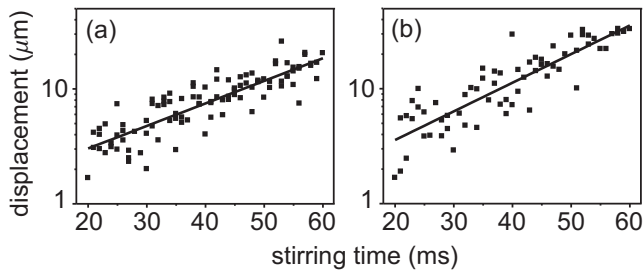}}
 \caption{Center-of-mass
displacement after free expansion (log-scale) \emph{vs.}~stirring
time for $\Omega\!=\!\omega_\bot$ and $\epsilon\!=\!0.09$. (a)
Non-condensed cloud with $10^7$ atoms, $T\!=\!5\;\mu$K; (b) Condensate
with $10^5$ atoms.
%, $T<50$~nK.
Solid line: exponential fit to the
data.} \label{fig:CMinstability}
\end{figure}

The center-of-mass displacement as a function of the stirring time
$t$ is shown in Fig.~\ref{fig:CMinstability}. We choose here
$\epsilon=0.09$ and $\Omega=\omega_\bot$, so that the motion
predicted by Eqs.(\ref{eq:Xmotion}-\ref{eq:Ymotion}) is
dynamically unstable. To ensure reliable initial conditions, we
deliberately offset the center of the rotating potential by a few
micrometers with respect to the atom cloud. We find the
instability for the center-of-mass motion both for the
non-condensed cloud (Fig.~\ref{fig:CMinstability}a) and for the
quasi-pure condensate (Fig.~\ref{fig:CMinstability}b). The
center-of-mass displacement increases exponentially, with an
exponent consistent with the measured $\epsilon$.

\begin{figure}
\centerline{\includegraphics{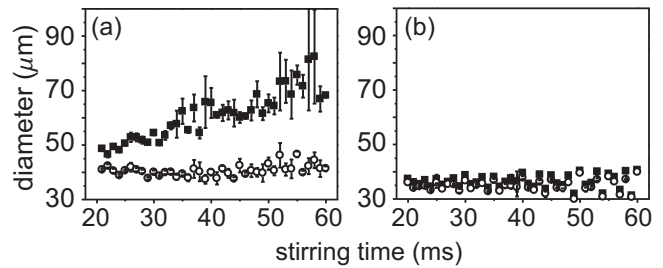}}
\caption{Long {\small($\blacksquare$)} and short ($\circ$)
diameters of the atom cloud
\emph{vs.}~stirring time, for $\Omega=\omega_\bot$. (a)
Non-condensed cloud~; (b) Quasi-pure condensate (same parameters
as in Fig. \ref{fig:CMinstability}).}
\label{fig:sizeincrease}
\end{figure}

We consider now the evolution of the size of the atom cloud as a
function of $t$ (Fig.~\ref{fig:sizeincrease}). The non-condensed
cloud exhibits the behavior expected from the single particle
dynamics, \emph{i.e.} the ``explosion'' in the $X=Y$ direction.
The cloud becomes more and more elliptical in the $xy$ plane. The
long radius increases with time, while the short one remains
approximately constant (Fig.~\ref{fig:sizeincrease}a). On the
opposite, we find that the condensate remains circular
(Fig.~\ref{fig:sizeincrease}b), with no systematic increase in
size. We then obtain the following counter-intuitive result: for a
significant repulsive interaction the atoms remain in a compact
cloud, while they fly apart if the interaction is negligible. We
observe this stability of the shape of the condensate rotating at
the critical velocity for $\epsilon \leq 0.2$. Above this value of
$\epsilon$ we find that the atomic cloud rapidly disintegrates.
For $\epsilon\approx 0.3$, after a stirring time of $50$~ms, we
observe several fragments in the time-of-flight picture.

We now perform a theoretical analysis of how the interparticle
interaction stabilizes a rotating condensate. To this end we use
the 2D ($x,y$) time-dependent Gross-Pitaevskii (GP) equation for
an idealized cylindrical trap ($\omega_z=0$). In the rotating
frame the GP equation reads:
\begin{equation}
i\partial_{t}\psi\!=\!\frac{1}{2}\left[\!-\Delta\!
+\!(\!1\!-\!\epsilon)X^{2}\!\!+\!(\!1\!+\!\epsilon)Y^{2}\!\!
+\!2g|\psi|^{2}\!\!-\!2\Omega\hat{L} \right]\psi,\!\!
\label{eq:gpe}
\end{equation}
where $\hat L$ is the $z$ component of the angular momentum
operator. In Eq.(\ref{eq:gpe}) the coordinates are given in units
of the initial harmonic oscillator length
$\sqrt{\hbar/m\omega_\bot}$, and the frequencies in units of
$\omega_\bot$. The condensate wave function $\psi(X,Y,t)$ is
normalized to unity, and the effective coupling constant is
$g\!=\!4\pi a\tilde N $, with $a$ being the positive scattering
length, and $\tilde N$ the number of particles per unit axial
length. The effective coupling $g$ depends on density and
characterizes the  ratio of the mean-field interparticle
interaction to the radial frequency $\omega_\bot$.

Since the trapping potential is harmonic, the average center of mass
motion of the condensate is described by the classical equations
(\ref{eq:Xmotion}-\ref{eq:Ymotion}) and is decoupled from the
evolution of the condensate wave function in the center of mass
reference frame. We shall therefore restrict to wave functions
$\psi$ centered at $x=y=0$ for all times.

We start with a variational analysis of the steady state of the
condensate in the rotating frame, using a Gaussian ansatz for the
condensate wave function \cite{cirac}:
\begin{equation}
\psi(X,Y) \propto \exp(i\alpha XY-\beta X^2/2-\gamma Y^2/2)\ .
\end{equation}
We use the symmetry properties of the Hamiltonian and assume that
the condensate wave function remains invariant under the
combination of a time reversal ($\psi \rightarrow \psi^{*}$) and a
reflection with respect to the $XZ$ plane. This implies that the
parameters $\alpha$, $\beta$, $\gamma$ are real numbers. We
extremize the GP energy functional with respect to these
parameters. Extremizing with respect to the phase parameter
$\alpha$ gives $\alpha=\Omega(\gamma-\beta)/(\gamma+\beta)$. As
$\beta$ and $\gamma$ should be finite and positive this sets the
constraint $\alpha^2<\Omega^2$. Extremizing over $\beta$ and
$\gamma$ and expressing $\beta/\gamma$ in terms of $\alpha$, we
obtain a closed equation for $\alpha$:
\begin{eqnarray}
\!\!\!& & (\epsilon/\Omega)\,[\alpha^2+2\alpha\Omega(\Omega^2-1)
/\epsilon+\Omega^2]    \nonumber\\
\!\!\!& & -(g/2
\pi)\sqrt{1-\alpha^2/\Omega^2}\,[\alpha^3+(1-2\Omega^2)
\alpha-\Omega\epsilon]=0.\,\,\,\,\,\,\,\,\,\, \label{eq:great}
\end{eqnarray}
In the non-interacting case ($g=0$) the Gaussian ansatz is exact,
and $\alpha$ is the root of a quadratic equation. This ansatz also
captures the scaling properties of the rotating condensate in the
regime of strong interactions. In the TF limit ($g\rightarrow
\infty$) the first line of (\ref{eq:great}) can be neglected and
we recover the cubic equation for $\alpha$ derived in
\cite{Stringari}.

For $g=0$ the parameter $\alpha$ is complex in the interval of
rotation frequencies $\sqrt{1 - \epsilon}<\Omega<\sqrt{1 +
\epsilon}$, and there is no steady state solution for the
condensate wave function at these $\Omega$ \cite{Stringari}. For a
finite $g$ the lower border $\Omega_-$ of this frequency interval
remains equal to $\sqrt{1-\epsilon}$ irrespective of the value of
$g$. The upper border, dashed curves in Fig.3, decreases with
increasing $g$ at a given $\epsilon$. For small anisotropy
$\epsilon<1/5$ it reaches the lower border at a critical coupling
strength. For larger $g$ the steady states exist at any $\Omega$.
If $\epsilon>1/5$, the upper border never reaches
$\Omega_-=\sqrt{1-\epsilon}$ and for any $g$ one has an interval
of $\Omega$ where steady state solutions are absent. The
$\epsilon=1/5$ threshold was derived for the TF limit in
\cite{Stringari}.

We now analyze the time evolution of the condensate after the
stirring potential has been switched on. For this purpose we use
an approximate scaling approach to the solution of
Eq.(\ref{eq:gpe}). We assume (and later on check) that the
evolution of the condensate shape is well described by dilations
with factors $b_{u}(t)$ and $b_{v}(t)$ along the axes $\tilde X$
and $\tilde Y$, rotating at an angular frequency $\dot\phi(t)$
with respect to the laboratory frame $x,y$. To determine $b_{u}(t),
b_{v}(t)$, and $\phi(t)$, we write the wave function as
\begin{equation}
\psi(\tilde X,\tilde Y,t)=(b_{u}b_{v})^{-1/2}
\chi(u,v,t)\exp\{i\Phi(\tilde X,\tilde Y,t)\} \ ,
\label{eq:scaling}
\end{equation}
where we have set $u=\tilde X/b_{u}$, $v=\tilde Y/b_{v}$, and
\begin{equation}
\!\Phi(\tilde X,\tilde Y,t)={\tilde \alpha}(t)\tilde X\tilde
Y\!+(\dot b_{u}/2b_{u})\tilde X^2\!+(\dot b_{v}/2b_{v})\tilde
Y^2\!,
\end{equation}
with ${\tilde \alpha}=-\dot \phi \tanh\xi$ and
$\xi(t)=\ln(b_{v}/b_{u})$. Then the GP equation reduces to the
following equation for $\chi(u,v,t)$:
\begin{eqnarray}
& & i\left(\partial_t -\frac{\dot
\phi}{\cosh\xi}(u\partial_v-v\partial_u) \right)\chi =
\Big[-\frac{\partial^2_u}{2b_{u}^2}-\frac{\partial^2_v} {2b_{v}^2}
\nonumber \\
& & \qquad \quad + \frac{1}{2}\left(\nu_{u}^2b_{u}^2u^2
+\nu_{v}^2b_{v}^2v^2 \right)
+\frac{g|\chi|^2}{b_{u}b_{v}}\Big]\;\chi. \label{rescaledGP}
\end{eqnarray}
The ``frequencies" $\nu_{u}$ and $\nu_{v}$ are given by
\begin{equation}\label{omegas}
\nu_{u,v}^2=1\mp\epsilon\cos(2\Omega t-2\phi)+{\tilde
\alpha}^2\mp 2\dot \phi {\tilde \alpha}+\ddot
b_{u,v}/b_{u,v}.
\end{equation}
In Eq.(\ref{rescaledGP}) we took into account that $u,v$ are
eigenaxes of the condensate, which requires the absence of terms
proportional to $uv$ and is provided by the relation
\begin{equation}
 \label{momentum}
\epsilon\sin(2\Omega t-2\phi)-\dot{\tilde \alpha}-{\tilde \alpha}
(\dot b_{u}/b_{u}+\dot b_{v}/b_{v})+ \dot\phi\dot\xi=0.
\end{equation}

We now replace the lhs of Eq.(\ref{rescaledGP}) by $\tilde \mu
\chi$, where $\tilde\mu$ follows from the normalization condition
for $\chi$. The solution of the resulting equation is a function
of $b_{u,v},\phi$ and $\nu_{u,v}$. We then require that
(\ref{eq:scaling}) is a relevant scaling transform, \emph{i.e.},
that the function $\chi(u,v,t)$ is most similar to the initial
function $\chi(u,v,0)$. More precisely we set the averages
$\langle u^2\rangle_t$, $\langle v^2\rangle_t$ equal to their
values at $t=0$. This fixes $\nu_{u}$ and $\nu_{v}$ in terms of
$b_{u},b_{v},\phi$. The solution of Eqs.(\ref{omegas}) and
(\ref{momentum}) then gives the desired scaling parameters.

The omitted lhs of Eq.(\ref{rescaledGP}) only insignificantly
changes $\nu_{u}$ and $\nu_{v}$. It vanishes in both the TF regime
and for an ideal-gas condensate. For the TF limit our procedure
gives the same results as the scaling approach of \cite{Olshanii}.

\begin{figure}
\epsfxsize=8.5cm \hspace{-0.5cm} \epsfbox{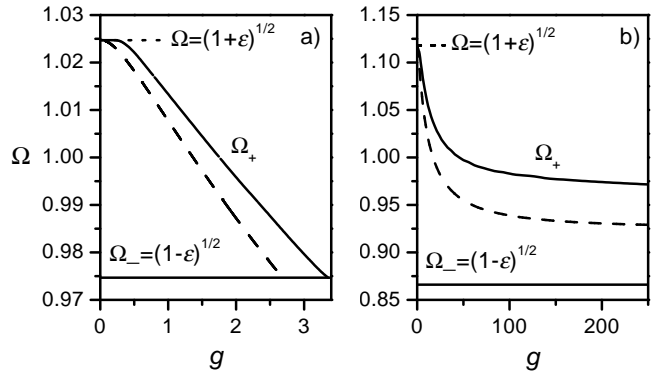} \vskip
-0.5cm \caption{\label{fig:delta1} Solid curves: upper
($\Omega_+$) and lower ($\Omega_-$) borders of the instability
region \emph{vs.}~$g$, for an abrupt switching of the rotation
with $\epsilon=0.05$ (a) and $\epsilon=0.25$ (b). Dashed curves:
upper border of the region where stationary states are absent.}
\label{fig:fig3}
\end{figure}

For an abrupt switching of the rotating potential we use the
initial conditions $b_{u,v}(0)=1$, $\dot b_{u,v}(0)=\phi(0)=\dot
\phi(0)=0$. We find two types of solution: (i) oscillating
functions $b_{u,v}(t)$, (ii) one of the scaling parameters
eventually grows exponentially. Case (ii) describes an infinite
expansion of the condensate in one direction, similarly to the
expansion of the ideal gas under rotation.

For a given $\epsilon$ we obtain the upper ($\Omega_+$) and lower
($\Omega_-$) instability borders in the $\Omega-g$ space (see
Fig.~\ref{fig:delta1}). The lower border is always equal to
$\sqrt{1-\epsilon}$. The upper border $\Omega_+(g)$ decreases with
increasing $g$ and for $\epsilon < 0.17$ it reaches $\Omega_-$ at
a critical value of the coupling strength. For $\epsilon> 0.17$ we
have $\Omega_+>\Omega_-$ at any $g$.

The obtained results can be understood by comparing the frequency
$\Omega_q$ of the rotating quadrupole mode of the condensate with
the rotation frequency $\Omega\sim 1$ of the perturbation
$V_1({\bf r})$. In the absence of interaction one has
$\Omega_{q}=1$, and the corresponding resonance leads to the
condensate explosion. The interactions reduce the frequency of the
rotating quadrupole mode ($\Omega_q=1/\sqrt{2}$ in the TF limit),
suppressing the resonance at $\Omega\approx 1$: the deformation of
the condensate induced by $V_1$ remains small, at least for
$\epsilon$ smaller than the detuning from the resonance. For
larger $\epsilon$ the condensate explodes.

In the presence of interactions ($1/\sqrt{2}<\Omega_q<1$), one could
expect naively that the explosion occurs for a resonant drive with
$\Omega \sim \Omega_q$, even for small $\epsilon$. This is not the
case because of a non-linear character of the dynamics. As the
system starts to elongate under the action of the resonant
excitation, it becomes closer to an ideal gas, for which
$\Omega_q=1$. The gas is then driven away from the resonance and
its deformation stops. This explains why the lower
instability border $\Omega_-$ is independent of $g$.

The scaling method can also be used to identify stationary
solutions, by setting $\dot \phi=\Omega $ and constant $b_{u,v}$
in Eqs.(\ref{omegas}-\ref{momentum}). The results nearly coincide
with the ones from the Gaussian ansatz. The existence of
these solutions can also be explored using an adiabatic switching
of the rotating potential. As shown in Fig.~\ref{fig:delta1} the
domain of instability for an abrupt switching of the rotation
includes the domain for the absence of stationary solutions.

\begin{figure}
\epsfxsize=8.5cm \hspace{-0.5cm} \epsfbox{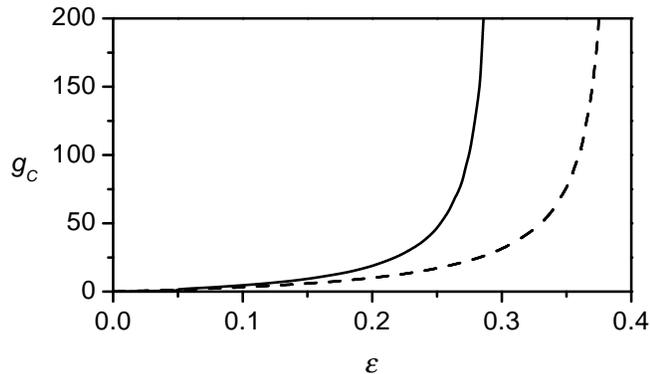} \vskip -0.5cm
\caption{\protect Critical coupling strength $g_c$ \emph{vs.}
$\epsilon$ for $\Omega=1$. The solid and dashed curves correspond
to an abrupt and adiabatic switching of the rotation,
respectively.} \label{PSh}
\end{figure}

In Fig.\ref{PSh} we present the minimum coupling strength
$g_c(\epsilon)$ required for the stability of the shape of the
condensate, for both abrupt and adiabatic switching of the
potential rotating at $\Omega=1$.

To summarize, our analysis shows that the condensate can preserve
its shape and size for rotation frequencies in the instability
window for the center of mass motion,
$\sqrt{1-\epsilon}<\Omega<\sqrt{1+\epsilon}$. This means that the
condensate will spiral out of the trap as a rigid body after the
rotation is switched on. For TF condensates this is the case if
$\epsilon\alt 0.28$, which explains why the repulsive
interparticle interaction maintains particles together in our
experiment. On the other hand, for larger $\epsilon$ there are
always frequencies at which even TF condensates are unstable. This
gives an account for the destruction of the condensate in our
experiment at $\epsilon \approx 0.3$. One can think of observing
related effects in rotating ion clouds in electromagnetic traps
(see \cite{Bollinger} and refs. therein).

Even though our picture properly describes the experiment, we are
likely not to deal with the ground state of the system. On a very
long time scale, the rotating gas can evolve to a more complex
state, for instance to a multi-vortex
state (with possible quantum melting) discussed in
\cite{Rokhsar,Ho,Gunn,Fetter,Baym,Sinova}, or to a Quantum-Hall-like state
\cite{Gunn,Zoller}. However these states have been
discussed for the axially symmetric case and a natural development
will be to include a finite rotating anisotropy $\epsilon$, which
is a necessary ingredient for most present experiments. In
particular, for $\Omega=\omega_\bot \sqrt{1- \epsilon}$, one
reaches a 1-body Hamiltonian corresponding to an unbound motion
(with a gauge field) in the $X$ direction, similar to a Quantum Hall
fluid in a narrow channel. We believe that the study of many body
aspects of this regime will bring in new features of quantum
mesoscopic physics.

We thank K. Madison for his help in early experiments and we
acknowledge fruitful discussions with J. Bollinger, S. Stringari,
and D. Wineland. P. R. acknowledges support by the Alexander von
Humboldt-Stiftung and by the EU, contract no. HPMF CT 2000 00830.
D.P. and G.V. acknowledge support from the Dutch Foundations NWO
and FOM, and from the Russian Foundation for Basic Research. This
work was partially supported by the R\'egion Ile de France, CNRS,
Coll\`{e}ge de France, DRED and INTASs.

\end{document}